# Enabling a Zero Trust Architecture in a 5G-enabled Smart Grid

Mohammad Ali Alipour [a], Saeid Ghasemshirazi [b], Ghazaleh Shirvani [c]

[a] Department of Energy Management and Optimization, Institute of Science and High Technology and Environmental Science, Graduate University of Advanced Technology, Kerman, Iran, m.alialipour77@gmail.com

[b] Department of Industrial Engineering, Iran University of Science and Technology, Tehran, Iran, saeidgs@yahoo.com

[c] Department of Computer Engineering, Iran University of Science and Technology, Tehran, Iran, ghazaleh.sh3p@gmail.com

**Abstract**

One of the most promising applications of the IoT is the Smart Grid (SG). Integrating SG's data communications network into the power grid allows for gathering and analyzing information from power lines, distribution power stations, and end users. A smart grid (SG) requires a fast and dependable connection to provide real-time monitoring through the IoT. Hence 5G could be considered a catalyst for upgrading the existing power grid systems. Nonetheless, the additional attack surface of information infrastructure has been brought about by the widespread adoption of ubiquitous connectivity in 5G, to which the typical information security system in the smart grid cannot respond promptly. Therefore, guaranteeing the Privacy and Security of a network in a threatening, ever-changing environment requires groundbreaking architectures that go well beyond the limitations of traditional, static security measures. With "Continuous Identity Authentication and Dynamic Access Control" as its foundation, this article analyzes the Zero Trust (ZT) architecture specific to the power system of IoT and uses that knowledge to develop a security protection architecture.

1. **Introduction**

Electricity is crucial to our contemporary way of life and economy; however, most nations still use power grids that were put in place around 50 years ago, making them inefficient and unable to meet the urgent needs of the modern world[1]. An estimated thirteen trillion dollars investment will be needed to upgrade our energy infrastructure over the next two decades to meet the demands of a modern, efficient power grid. As a result, there is a looming need and opportunity to switch to a low-carbon, efficient, clean energy system.

Smart grids will be an essential facilitator of this revolution. SG is an intelligent digitized energy network, optimally delivering electricity from generation to consumption. Integration of data, communications, and power technologies into the grid makes this possible. Some advantages of a smart grid are:

- Increased productivity and sustainability of energy generation
- More renewable energy sources will be incorporated into the preexisting infrastructure
- Supporting the development of electric vehicles
- Innovative strategies to help consumers reduce their energy consumption
- Less carbon dioxide emissions

Smart Grid is not just about enhancing the current infrastructure that helps our society; it's also about fulfilling the full potential of what we can provide, such as innovative transportation solutions, support for new markets, and the most efficient use of available resources. To provide constant connectivity between power companies and their consumers, a Smart Grid implements technologies that combine computing and networking with physical mechanisms. However, physical issues and cyber-attacks continuously threaten the availability and integrity of the grid, which might have negative social and financial consequences. Natural disasters and reliance on the public internet, which is susceptible to cyberattacks, are increasing the frequency of malfunctions and breaches[2].

Reliability of the electricity system is essential to guarantee a high standard of living. Because of its vital importance, the power system is vulnerable to cyberattacks that might spread panic and have far-reaching financial impacts. A great example of this could be Ukraine's electricity infrastructure cyberattack in 2015[3]. Consequently, an unauthorized intrusion caused an outage in the distribution firms' SCADA system.

Many different types of attacks apply to this zone; some examples are replay attacks, Dos attacks, MITM attacks, and false data injection attacks[4]. For a more reliable smart grid, this article suggests a policy that characterizes "Continuous Identity Authentication and Dynamic Access Control", known as the Zero Trust approach[5,6]. Nevertheless, Zero Trust at the IoT or smart grid scale is impractical due to the curated communication rate and bottlenecks. As research delves into zero trust for IoT, they indicated that one major factor that promises a practical application of ZT is 5G. So far as we are aware, this research is the first to explore 5G applications utilizing ZT to improve smart grid security.

Fifth-generation wireless networks incorporating a 5G network slicing approach could offer smart grid services, including grid monitoring, advanced metering infrastructure (AMI), and precise load control[7,8]. 5G connectivity features high bandwidth, low latency, excellent dependability, and low power usage. Therefore, 5G technologies have the potential for novel applications due to their enhanced mobile capacity, ultra-low latency communication, and universal terminal access. With the Utilization of 5G communication, it would be possible to streamline the gathering and analysis of data on power usage and enhance the precision with which power loads are managed[9,10].

The perception and transmission performance of numerous user nodes in the Internet of Things can be improved by developing a 5G cognitive radio network model and its application to conventional gathering and inspection services IoT. The ability to acquire and visualize data for various smart grid levels is a potential advantage of 5G technology for the future smart grid. Smart grid applications bring new vulnerabilities, such as security misconfiguration at edge hosts and IoT device security concerns, and most importantly that 5G networks cannot offer end-to-end Security for these scenarios[11].

When using wireless technologies like 5G for mission-critical applications in the smart grid, further precautions must be taken to prevent unwanted access to the network. Therefore, numerous 5G smart grid AMI components might be subject to denial-of-service (DoS) or fake data injection attacks, leading to financial or nonfinancial repercussions[12,13].

## 2. Contributions

This paper's primary goal is to apply the Zero Trust mindset to a smart grid with a 5G communication network. This security model was neither employed nor applied to the communication network in the previous papers. In this paper, we apply Zero Trust in 5G-enabled Smart Grid To cover this gap. The following could serve as a summary of the main contributions: A comprehensive review of the smart grid and zero trust was carried out, and each component was explained. Moreover, in the methodology section, the communication network and smart grid components employing zero trust, such as breakers, are described. Consequently, zero trust is applied to the modeled 5g network. At the time of writing this paper, no study assessing zero trust within the 5G-enabled Smart Grid exists.

## 3. Background

### 3.1. Smart Grids

The National Institute of Standards and Technology (NIST) describes Smart Grid as consisting of seven subsystems: generation, transmission, distribution, markets, customers, service providers, and operations. The end-user is the primary participant in the customer domain. Customers typically fall into three categories: residential, industrial, and building or commercial. In addition, some actors are prosumers, which means they can consume electricity and generate it at the same time[10]. This domain is electrically linked to the distribution area and communicates with the distribution, marketplaces, service provider, and operation[14].

Electrical power is transmitted from the generating domain to the distribution domain through a series of substations across great distances. Additionally, it can be used to generate and store electricity. The SCADA system used to monitor and operate the transmission network includes communications, monitoring, and controlling tasks[15].

The term "distribution" describes the whole ecosystem of companies and organizations that move energy from generation to consumption. Radial, looping, and mesh topologies are only a few possible arrangements for an electrical distribution network. Energy production and storage are included in the distribution domains linked to the transmission system, the ultimate customer, and power consumption metering stations.

Electricity market operators and consumers are the actors in this sector[16]. This space keeps up the harmony between the supply and request of power. Consequently, To adjust supply and demand, the showcase interatomic with vitality supply spaces, such as the spaces for Bulk Generation and dispersed vitality assets[10].

### 3.2. Zero Trust

Regarding information security, the Zero Trust paradigm stipulates that no one, within or outside the network, can be trusted. The "never trust, always verify" philosophy dictates that businesses must take precautions to protect the confidentiality of any data stored on any device accessible by any user, application, or network[17].

When it comes to Security, ZT isn't concerned with securing individual nodes but rather the resources those nodes contain; in other words, instead of assuming that every user on a network is trustworthy and has permission to access its resources, ZT assumes that nobody on the network, internal or external, has either. This is a direct response to growing needs inside businesses for securing their cloud-based assets and remote workers[18].

Therefore, ZT perception might signal a revolutionary change away from "trust but verify" to instead never trust and always verify. Here are the fundamentals of zero trust (ZT), as outlined by the US National Institute of Standards and Technology Publication 800-27[19]:

- **Integrity Monitoring:** Devices' security status and user/network asset activity patterns are thoroughly evaluated by a system.

- **Least Privilege**: The "least privilege" principle dictates that the lowest possible level of access permissions should be used wherever possible. Depending on the sensitivity of the source, access has only been provided to that resource and will not work for any others.

- **Dynamic Policy:** The decision of whether or not to give access requires a dynamic policy. The subject and network assets' behavioral characteristics and security status, including credentials, software patches, location, and the like, are crucial choice variables.

- **Trust Evaluation:** All access requests undergo a thorough trust analysis and risk review. The evaluation is ongoing (during the access time) and variable (based on situational conditions).

### 3.3. 5G Technology in Smart Grids

Many electrical items are connected to smart grids for enormous monitoring and connectivity. Their interconnectedness will give smart grids limitless creative possibilities. By integrating and coordinating all home appliances, for instance, the power consumption of buildings, towns, or even an entire area may be dynamically changed to accommodate distributed renewable energy[7,20]. 5G enables mobile devices to connect with excellent dependability. Through V2G charging stations, the connectivity of electric cars contributes to smart cities and smart transportation and helps the power grid accept distributed renewable energy[21].

Due to the communication overhead, even though many sensors have been deployed, enormous amounts of data must be deleted, and only the most essential information may be saved. In smart grids, the absence of granular data severely hampers the practical implementation of big data analytics. Other monitoring devices, such as micro-PMU, will be deployed in the future. 5G connection enables the processing of diverse and massive amounts of data. Even if the applications of 5G communication in smart grids are promising, several obstacles, such as high energy consumption, coverage, and resource distribution among various businesses, must be resolved in the future.

4. **Methodology**

   4.1. **Smart Grid Architecture**

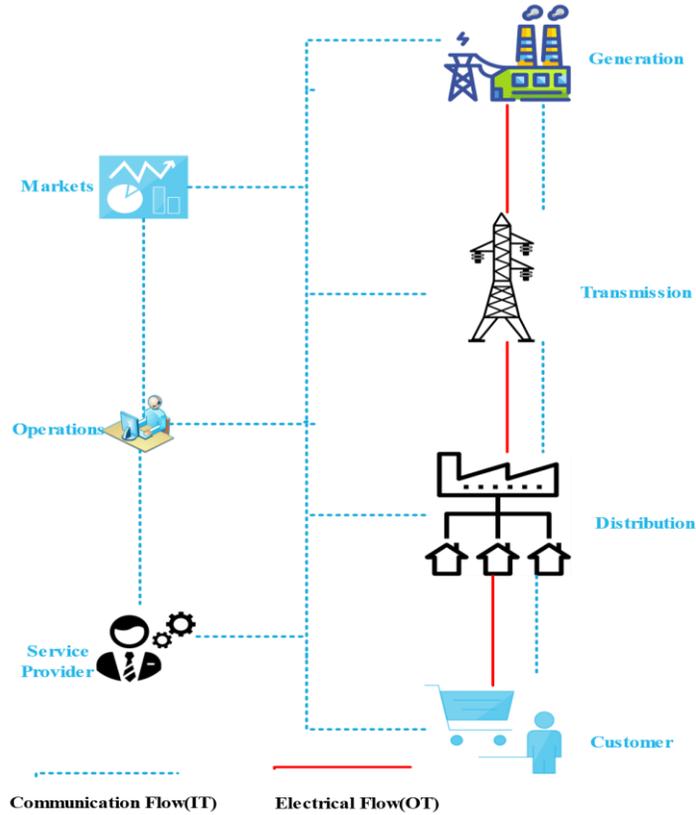

**Figure 1. Most important smart grid domains by NIST**

The smart grid may be classified into three sections, as seen in Figure 2, and each is mainly analyzed, as seen in the above chart.

1. The physical layer of the power system: includes power distribution, transmission, and generation utilities.

2. The data transport and control layer involves information transfer and systems management. A high-speed, all-inclusive communication network accelerates data collection, and secure data sharing enables connectivity between the parties.
3. Application layer: automatic metering, maintenance planning programs, and end-user broadband access[22].

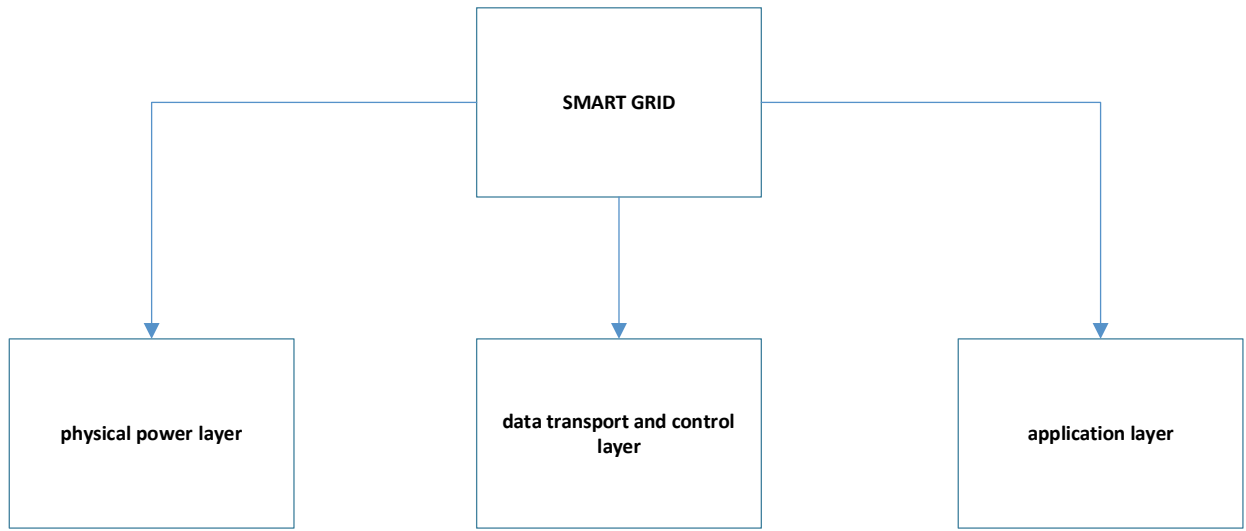

**Figure 2. Smart Grid Classification Scheme**

AMI and communication networks [10] provide the basis of the smart grid by storing, transmitting, and processing data in smart grids. As all smart grids have a great connection with the market, and this equipment can also run a command such as cut-off power, etc., this communication should be secure, fast, and reliable. The main work of this study is the data transport and control layer.

In addition, smart meter technologies provide smart control of devices to analyze their performance and give instantaneous system reactions. Smart Meters and Phasor Measurement Units (PMU) are unavoidable among the components necessary for advanced Smart Grid functioning. The unique aspect of the infrastructure for autonomous measurement and data collecting is the use of real-time rates based on demand. This work is the most significant since gathering real-time data on power load will aid in establishing the power price to achieve load-generation balance[23,24]. Using cellular technology, AMI offers two-way communication between SGs and MDMS (meter data management devices)[25].

After knowing the significance of communication in the SG, we will propose a new approach for Security and this network. The proposed method is based on a zero-trust approach, and we will work on policies and rules in every section of our smart grid communication network.

Available wireless networks do not have the reliability that all smart meters data in the network deliver to the MDMS unit at a suitable time[26]. For this study, the 5g network was selected as the optimal channel for smart grid communications. Several countries throughout the globe are actively pushing for the adoption of fifth-generation (5G) cellular network technology due to its numerous advantages over older generations in areas such as data transfer rate, network uptime, safety, energy efficiency, and connection number[7,27,28]. The electromagnetic wave frequency range of 5G could extend into the dozens and dozens of gigahertz, much beyond 1G–4G[29]. Therefore, the 5G wavelength is shorter, and the data transmission bandwidth is wider. Experiments by[30] show that 5G networks can ensure radio transmission in the sub-millisecond range. In the following discussion, we'll examine the 5G's core features that make it an ideal option for a smart grid.

- **Millimeter wave (mmWave):**

    5G has the highest wave frequency between 3.4 GHz and 3.8 GHz and 24 GHz, and 100 GHz.at 100 GHz, the wavelength of 5G can exceed 3 millimeters. Consequently, the high-band wavelength used by 5G is also known as mmWave or millimeter wave[31]. The shorter wavelengths of mmWave allow for the creation of tighter beams, allowing mmWave to transfer data faster, with reduced delay, more resolution, and more privacy.

- **Massive multiple-input multiple-output (MIMO):**

    Since the low-band spectrum's components are more important, wireless devices are limited to fewer antennas than they would be able to use with mmWave (high-band) components. Massive multiple input and output (MIMO) is the term for this innovation[31]. 5G gadgets typically include 32–256 antennae.

- **Ultra-dense cellular networks:**

    In addition to increasing data transfer speed and Security, compared to mmWaves, which can travel hundreds of meters, 5G's transmission range is limited by its shorter wavelength and tighter beams. Furthermore, mmWaves are easily obstructed by objects such as walls, vegetation, and human bodies. For 5G cellular networks to provide

seamless coverage, a greater number of 5G tiny cells, also known as an ultra-dense cellular network, must be constructed[31,32].

Existing research has categorized 5G wireless systems into three scenarios: improved mobile broadband (eMBB), ultra-reliable and low-latency communications (uRLLC), and massive machine-type communications (mMTC). Using a 5G network, one million devices can be connected per square kilometer. Thus, it's possible to use terminal or embedded controllers to manage the networked appliances.

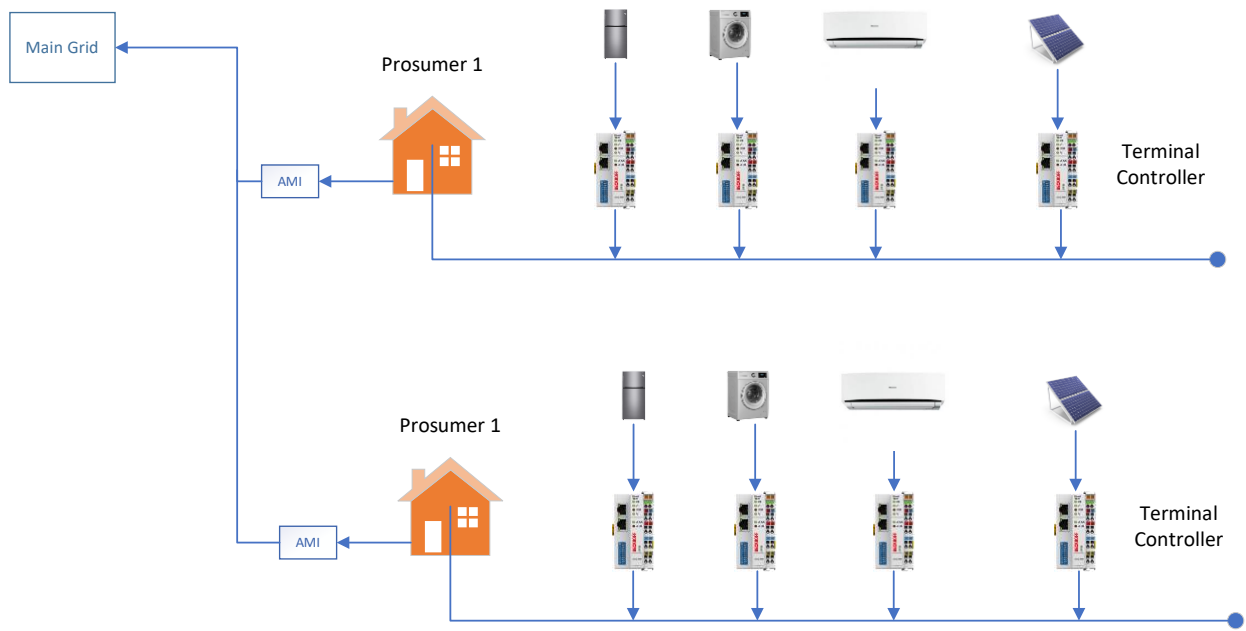

**Figure 3. 5G's mMTC (massive machine type communication) application in smart grid to make connections between numerous devices.**

Four data channels at one data frame every 15 or 30 minutes must be stored in AMI for at least 45 days. AMI needs mMTC technology since millions of smart meters are already installed in current distribution networks. As seen in Figure 4, 5G's mMTC (massive machine type communication) feature allows for a large number of connections between various loads, improving DR (demand-response) management and grid monitoring[33].

PMUs are required for Smart Grid's long-distance monitoring, control, safeguarding, and self-healing functions[34]. Delays of 20-50 milliseconds are typical for 50-hertz electricity networks, and the required PMU reporting rate can exceed 50 data frames per second. Due to high investment costs, distribution networks won't have many PMUs. PMUs will only be utilized for the most crucial Smart Grid operational functions and put on the most critical buses. Reporting rate, latency, the quantity of PMUs, and high-reliability requirements for protection and control show that uRLLC is needed. mMTC and uRLLC must be used simultaneously in Smart Grid. mMTC and URLLC should be supported for synchronous PMU and AMI applications over 5G[20].

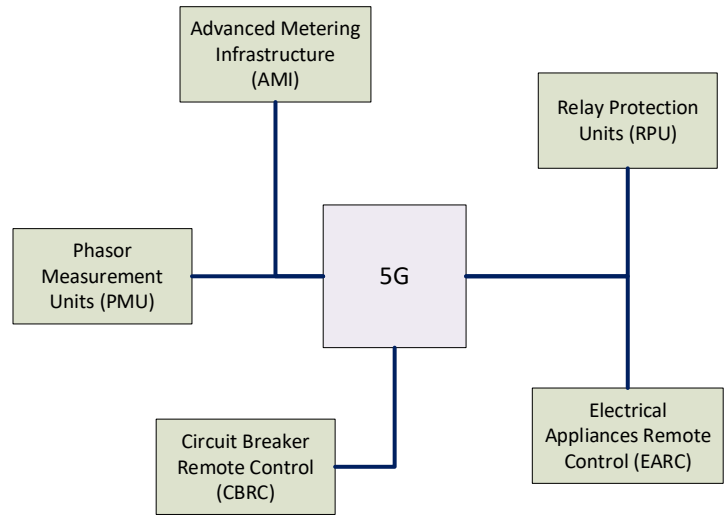

Figure 4. depicts typical distribution system components that could be linked to the future 5G network. It is anticipated that remote control of circuit breakers may shorten the duration of power outages. To execute demand-side management services, distribution system operators might remote-control electrical appliances. The sensitivity, selectivity, and tripping time of relay protection devices might be enhanced by mutual communication.

Because of these enhancements, 5G is becoming the preferred method of communication for smart grids. While 5G networks can improve data security through network slicing, SDN, and NFV, there are more potential security risks than wireless and private optical fiber communication networks. Existing private networks for smart grids can achieve physical isolation with the extranet. Several studies[35,36] have shown that.

### 4.2. ZT for 5G Grid

5G wireless networks are anticipated to manage a significant amount of data created by various devices, such as smartphones, autonomous cars, smart buildings, and smart grids. Smart grids, which deal with national Security and public safety, heavily rely on network-based data management and processing[28,37].

The authentication process in conventional cybersecurity models builds trust in a network's ability to provide a user or device access to protected resources, including data, apps, and services. The moniker "zero trust architecture" (ZTA) comes from the fact that authentication in and of itself does not establish trust. The Zero Trust Architecture requires authentication instead of blind trust before granting access[38,39].

### 4.3. Zero-Trust Model for 5G Service Providers

Communication service providers (CSPs) in the telecoms business require a robust strategy to protect their infrastructures from known security threats. A conventional protection technique first targets central routing services at the network's perimeter. With a zero-trust approach, Security may be taken to the next level. With zero trust, neither inside nor outside entities are trusted; dynamic policies govern resource access. Regarding policy decision-making and boosting security posture, zero trust depends on the integrity, behavioral monitoring, and security analytics. Researchers advise CSPs to implement a zero-trust policy and utilize advanced analytics to increase the level of security protection. A security platform must have a solid base. Moreover, the operators of 5G networks demand a robust and comprehensive security policy that incorporates all traffic across the signaling, data, and application levels. While implementing 5G security and implementing Zero Trust principles, service providers have several opportunities to improve Security:

1. **Least Privilege:** Use micro-segmentation to more precisely protect 5G network services.

2. **Precise Security Policies:** Business clients implement more focused security policies for finer-grained control over data and application access.

3. **Protect Workloads Utilizing Cloud-Native Network Function (CNF):** Preserve CNF throughout the lifecycle of CI/CD.

4. **Identify and Prevent Threats:** Monitor any interactions between network layer functions.

5. **Utilize Automation and Artificial Intelligence:** Improve anomaly detection to prevent distributed denial of service (DDOS) attacks.

6. **Implement Security using Access Requests:** Focus less on securing the entirety of the attack surface.

In essence, Zero Trust for 5G allows service providers, corporations, and organizations to reconsider how users, apps, and infrastructure are secured in a sustainable and maintainable manner for modern cloud, SDN-based environments, and open 5G networks. Enabling the Zero Trust Enterprise necessitates implementing Zero Trust principles and revamping security procedures to keep pace with digital disruption.

## 5. Discussion

Digitalization has numerous positive effects on the power grid but also increases cybersecurity threats. A successful cyberattack might lead to losing control over equipment and operations, leading to significant service interruption and possibly even physical harm. As reported by the French think-tank Institut français des relations internationales (IFRI), the number of cyber-attacks on the electricity sector increased by a whopping 380% between 2014 and 2015. Power Technology explores the five most significant cyberattacks against the power sector in the past few years. There are a variety of possible motivations, including geopolitical, sabotage, and financial targets. Some of the cyberattacks on the power system are listed here:

1. In western Ukraine, hackers entered the system of a power company, cutting power to 225,000 homes. (December 2015)
2. After destroying a power substation, hackers in Kyiv left the city's residents in the dark for one hour. According to the BBC, the blackout caused Kyiv to lose enough electricity to run its lights for an entire evening. Some researchers believe the attack was intended to cause actual physical damage to the power grid. (December 2016)
3. The United States and Israel created the computer worm Stuxnet to target Iran's nuclear facilities. It disabled safety mechanisms, resulting in uncontrollable centrifuges. Microsoft Windows is utilized by Stuxnet to target Siemens industrial control systems. Hackers have previously attacked industrial systems, but this is the first malware that spies on and subverts them, including a PLC rootkit. (June 2010)

We need a method with high reliability to safeguard the electrical system against cyberattacks such as those described above. In prior research, numerous security methods and techniques were applied to the smart network; however, applying zero trust to the 5G platform and smart networks is a novel approach. This article discusses the modeling and prerequisites of the smart grid that operates on the 5g network, the capabilities of this platform, and how to implement the rules that meet the zero trust standards. Few studies have addressed zero trust and smart grids, but these studies have addressed this topic. However, none of these studies[40,41] has addressed enabling zero trust security rules for the smart grid's 5G network. Finally, this article presents a comprehensive strategy for developing a smart and secure grid by combining three modern technologies.

## 6. Conclusion

The existing security solutions cannot manage and safeguard the large scale of connected grids in the 5G arena. Given the continuously changing threat landscape, the traditional strategy of "trust but verify" does not meet the security needs of vital infrastructures, such as the smart grid system. Integrating new 5G communications and networks with future Smart Grids is a primary goal of the 5G Grids, which holds great potential for boosting future Security and trustworthiness through cutting-edge services. With the proliferation of access control and remote access methods, Smart grid operations are increasingly vulnerable to penetration and the propagation of cyberattacks; therefore, this research proposes a zero-trust approach for a Smart Grid with 5G connectivity.

Regarding protecting the 5G-enabled Smart grids, the zero-trust security model offers promising results in device authentication, least-privileged access, device health, continuous updates, security monitoring, and incident response. Moreover, when applied to grids, the ZT security paradigm effectively tracks and detects harmful actions from any user or piece of hardware connected to the network. By preventing and isolating threats using advanced security solutions that employ a zero-trust methodology, the frontline of defense of a Smart Grid can be enhanced.

## 7. Future Research

With the rapid growth of electric vehicles (EVs), the government's Net Zero Strategy suggests using vehicle-to-grid (V2G) technology before and after 2030. V2G feeds electricity back into the grid when supply exceeds demand and supports local grids and energy systems by allowing one-way energy exchange between a grid and a plug-in car. V2G has a good outline; nevertheless, it carries numerous privacy issues. V2G privacy requires hiding the vehicle's true identity and location from the Aggregator.

V2G's biggest privacy issue keeps the Aggregator from discovering the vehicle's true identity and location. The vehicle's identity may be compromised if the Aggregator shares critical information with an opponent. The Aggregator or operator may also record car-specific data like charging time and station location. The owner's schedule can determine whether a vehicle is charged or discharged at the same station. Therefore, zero-trust further study, akin to smart grids, might be a potential solution.

**Conflicts of Interest**

The authors declare that they have no conflicts of interest.